\documentclass[12pt]{article}
\usepackage[affil-it]{authblk}
\usepackage{amssymb,physics,slashed,graphicx,tikz}
\usepackage[margin=1in]{geometry}

\begin{document}
\title{$\eta_{\rm w}$-meson from topological properties of the electroweak vacuum}

\author{
\textsc{Gia Dvali$~^{1,2}$, Archil Kobakhidze$~^3$ and Otari Sakhelashvili$~^2$} 
\vspace{0.2cm} \\
\normalsize \itshape
$^1~$Arnold Sommerfeld Center, Ludwig-Maximilians-Universität,\\ \normalsize \itshape
Theresienstraße 37, 80333 München, Germany and \\ \normalsize \itshape
$^{2}~$Max-Planck-Institut für Physik, Boltzmannstr. 8, 85748 Garching, Germany \vspace{0.2cm} \\ 
\normalsize \itshape
$^{3}~$Sydney Consortium for Particle Physics and Cosmology, \\
\normalsize  \itshape
School of Physics, The University of Sydney, NSW 2006, Australia \\ 
}

\maketitle

\begin{abstract}

\noindent 

 We further scrutinize the evidence for a recently suggested pseudo-scalar particle, the electroweak $\eta_{\rm w}$-meson. Its existence is demanded by matching the removal of the weak vacuum angle $\theta_{\rm w}$ by the anomalous $B+L$ - symmetry with a massive pole in the topological susceptibility of the vacuum.  We specifically focus on the possibility of the emergence of $\eta_{\rm w}$ as a collective excitation of the phase of the condensate of the 't Hooft fermion determinant, generated by the electroweak instantons, which breaks the $B+L$ - symmetry spontaneously.  We argue that the generation of the 't Hooft vertex is in one-to-one correspondence with its non-zero vacuum expectation value which is  cutoff insensitive.  We outline certain puzzles about the nature of the emergent $\eta_{\rm w}$ which require further investigations. 
\noindent 

\end{abstract}

\maketitle

\section{Introduction} 

    Recently \cite{Dvali:2024zpc}, we have pointed out that the electroweak sector 
    of the Standard Model (SM) contains a new degree of freedom, the $\eta_{\rm w}$-meson. 
   The emergence of $\eta_{\rm w}$ is correlated with the 
   elimination of the electroweak vacuum angle 
   $\theta_{\rm w}$ by the anomalous $B+L$-symmetry 
   of the SM.   The connection between
   anomalous symmetry and nullification of 
   topological susceptibility  of the vacuum
   (TSV) demands the existence of a particle which 
   gets its mass from TSV \cite{Dvali:2005an}.   
   In the present context this particle is  sourced by the anomalous 
    $B+L$-current and gets its mass from the TSV of the electroweak gauge theory \cite{Dvali:2024zpc}. 
 In this sense the $\eta_{\rm w}$-meson of the electroweak theory plays the role analogous to $\eta'$-meson
 of QCD, which is sourced by the anomalous axial current and gets its mass from the TSV of QCD.

  Although the main indication for the existence of $\eta_{\rm w}$ comes from the gauge structure of the electroweak theory, for certain shortcut conclusions in \cite{Dvali:2024zpc} we used gravity as an external  
monitoring tool.  At the basic level 
 of reasoning leading to the existence of 
 $\eta_{\rm w}$, gravity was used as a spectator device without any assumptions beyond the properties defined by the ordinary general covariance. 
These features show that $\eta_{\rm w}$ is a 
necessary ingredient of the SM, the least, upon its coupling to gravity. 
This demand becomes even stronger within the $S$-matrix formulation of gravity, which is generically  inconsistent with  $\theta$-vacua
\cite{Dvali:2018dce, Dvali:2022fdv, Dvali:2023llt, Dvali:2024dlb}. 

 In the present paper, we wish to ignore gravity and focus on the evidence for $\theta_{\rm w}$
 provided by the electroweak sector of the SM. 
 The main question is whether $\eta_{\rm w}$ emerges as 
 the phase of the condensate of the fermion 
 't Hooft determinant. This order parameter 
 is generated by the electroweak instantons 
 and breaks the anomalous $B+L$-symmetry spontaneously.  We show that the non-zero vacuum expectation value (VEV) of the fermion determinant  is intrinsically linked with the 
 generation of the same structure  
 't Hooft vertex, which breaks $B+L$-symmetry explicitly.  As a byproduct of our analysis, 
 we clarify the issue about 
 the seeming  cutoff-sensitivity of the 
 fermion condensate that is exponentially growing in fermion  flavors. We argue that in the correct 
 treatment this sensitivity is absent, and the enhancement of the instanton amplitude 
 boils down to the exponential degeneracy factor 
 by the fermion zero modes. 

  At the level of our analysis, certain puzzles remain about the nature of the $\eta_{\rm w}$
  as of the collective excitation of the condensate phase. In particular, the relation between the validity domains of the effective field theoretic (EFT) descriptions of $\eta_{\rm w}$ and the SM fermions require a further clarification. 
   These are necessary for understanding whether 
 $\eta_{\rm w}$ is a collective excitation delivered  
by the electroweak sector or rather it represents 
a new elementary degree of freedom demanded by consistency.

   \section{Evidence from TSV and anomalous $B+L$-symmetry} 
   
  The main argument for the existence of  $\eta_{\rm w}$ 
presented in \cite{Dvali:2024zpc}  
  comes from applying to the Standard Model  the general correspondence developed in series of articles
  \cite{Dvali:2005an, Dvali:2005ws,  Dvali:2013cpa, 
Dvali:2017mpy, Dvali:2022fdv, Dvali:2023llt,Sakhelashvili:2021eid}.
  (for a detailed discussion and summary, see  \cite{Dvali:2017mpy}).
 
  This correspondence says that if a non-zero TSV  is made zero by a deformation of the theory, this deformation must produce a massive pseudo-scalar particle.   
   In particular, if the physics that makes TSV zero 
 is an anomalous $U(1)$ symmetry,  the theory must posses
 a pseudo-scalar particle
 shifting under this symmetry as well as
 a non-zero fermion condensate that breaks 
 $U(1)$ spontaneously. 
 
    In \cite{Dvali:2005an}  it was argued that the classic example 
   realizing this connection is provided by QCD with massless quarks. 
 In this case the anomalous symmetry that eliminates 
 TSV is the axial symmetry. The emerging particle is the 
 $\eta'$-meson, which is the phase of the chiral quark condensate and gets its mass from the instantons 
 \cite{tHooft:1976snw}.  
 
    It was further suggested that the same reasoning must apply to gravity \cite{Dvali:2005an}. Moreover it was argued 
   \cite{Dvali:2013cpa, Dvali:2016uhn, Dvali:2017mpy} 
     that in a theory with gravitational TSV and a chiral gravitational anomaly of fermions that kills it,  the fermion condensate must form, giving rise to a composite pseudo-scalar. 
   More recently \cite{Dvali:2024dlb}, we have clarified that due to the specifics of 
   gravitational Eguchi-Hanson instantons \cite{Eguchi:1978xp,Eguchi:1978gw}, the 
   fermions that eliminate TSV via gravitational anomaly must have spin-$3/2$ 
  (rather than spin-$1/2$) since the spin-$1/2$ fermions do not deposit the zero modes in the Eguchi-Hanson background. Furthermore, the same analysis restricts the fermion content in the presence of gravity, requiring cancellation of the total spin-$1/2$-gravitational anomaly \cite{Dvali:2024dlb}.
 
     In the present paper, we focus on the application of 
  this criterion to the electroweak sector of the SM 
  offered in \cite{Dvali:2024zpc} .   There we have argued that,
   despite the fact that the theory is in the Higgs phase,
   the correspondence is fully effective, leading to the existence of a new particle.  In particular, 
    the role of anomalous symmetry  that eliminates 
    TSV is played by $B+L$.    We have shown that the  $B+L$-violating fermion 
   condensate forms due to electroweak instantons. 
    Although, the instantons get constrained due to the Higgs 
    vacuum expectation value (VEV), the zero mode structure is effective since the fermion masses fully respect the $B+L$-symmetry. 
     We now discuss this in more details.

   \subsection{TSV in theory without fermions} 
   
   Let us consider the reduced version of the Standard Model from which we temporarily exclude all fermions. 
    We focus on the electroweak sector.  The 
    $SU(2)_W\times U(1)_Y$  gauge symmetry is in the Higgs phase.  This breaking however maintains the 
 topological structure of the vacuum, since the 
 instantons are not abolished but rather only become constrained. The $U(1)_Y$ part of the theory is irrelevant for the further discussion. 
  Correspondingly, the vacuum has an ordinary 
  $\theta$-vacuum structure \cite{Anselm:1992yz,Anselm:1993uj,Anselm:1990uy}. The choice of the vacuum is 
  accounted by  the boundary term, 
          \begin{eqnarray} \label{ST} 
  S_{\theta}\, &=&  \frac{\theta_{\rm w}}{16\pi^2}\int_{3+1}\, \,  W_{\mu\nu}\tilde{W}^{\mu\nu}\,, \nonumber  \\
  &&\,  W_{\mu\nu}\tilde{W}^{\mu\nu} \equiv \, \, \epsilon^{\mu\nu\alpha\beta}\, 
 {\rm tr} \, W_{\mu\nu} W_{\alpha\beta} \,,
\end{eqnarray}
where $W_{\mu\nu}$
 is the $SU(2)_W$ field-strength 
of the $2\times 2$ gauge field matrix $A_{\mu} \equiv A_{\mu}^c \tau^c$,  
with $\tau^c,~ c=1,2,3$ the three Pauli matrixes. 
 
 This term is a total derivative,  
  \begin{equation} \label{dCS}
  W \tilde{W} \,  = \, 
   \epsilon^{\mu\nu\alpha\beta} \partial_{\mu} C_{\nu\alpha\beta} \,,
   \end{equation}
 where, 
   \begin{equation} \label{CS}
   C_{\mu\nu\alpha}  \equiv {\rm tr} \left(A_{[\mu}\partial_{\nu}A_{\alpha]} + \frac{2}{3}A_{[\mu}A_{\nu}A_{\alpha]}\right) \,,
   \end{equation}
is the Chern-Simons  $3$-form.

     The physicality of the  $\theta_{\rm w}$-term  is directly linked      
   with the TSV correlator, 
  \begin{eqnarray}
    \label{EEcorr}
FT \langle W\tilde{W},  W\tilde{W}\rangle_{p\to 0} &&\equiv \\
\equiv \lim\limits_{p \to 0}\int d^4 x e^{ipx} \langle T [W\tilde{W}(x), W\tilde{W}(0)]\rangle &&=\mathrm{const} \nonumber\,,
\end{eqnarray}
where $T$ stands for time-ordering, $FT$ stands for  the Fourier transformation and  $p$ is a four-momentum. 
 The $\theta$-term is physical if the above correlator 
is non-zero and vice versa. 

We now come to the following key point. 
The expression (\ref{EEcorr})  implies that the 
  K\"all\'en-Lehmann spectral representation of the Chern-Simons correlator includes a physical pole 
    at $p^2=0$ \cite{Dvali:2005an, Dvali:2017mpy, Dvali:2022fdv, Dvali:2023llt},     
    \begin{equation}
    \label{CCcorr}
FT\langle C ,C \rangle =  \frac{1}{p^2} + \sum_{m\neq 0} 
\frac{\rho(m^2)}{p^2 - m^2} \,, 
  \end{equation} 
   where $\rho(m^2)$ is a properly normalized spectral function.  Notice that the presence of a massless 
pole $p^2 =0$ is a gauge invariant statement.  
    Since this pole appears in a $3$-form, it does not contain any propagating degrees of freedom. 
    
   Now, at non-zero coupling, the only gauge invariant way of making TSV zero is to shift the pole towards 
   a non-zero mass $p^2 =m^2 \neq 0$.  However, 
   a massive $3$-form propagates one pseudo-scalar degree of freedom.   
   This implies the 
   presence of a physical particle in the spectrum. 
   Thus, the same physics that renders the TSV zero, will bring the propagating particle in the spectrum of the theory.

    \subsection{$B+L$ - symmetry} 
    
     Next, let us restore the fermion content of the Standard Model.    The theory now
  has the global $U(1)_{B+L}$ symmetry, under which fermions
  transform as     
       \begin{eqnarray} \label{B+L}
   && (q_L, u_R, d_R)\, \rightarrow \, e^{i \frac16 \alpha } \,(q_L, u_R, d_R) \,,  \nonumber \\  
    && (\ell_L, e_R, \nu_R)  \,\rightarrow e^{i \frac{1}{2}\alpha} \, 
    (\ell_L, e_R, \nu_R) \,, 
\end{eqnarray}
 were $q_L \equiv (u_L, d_L)$ and $ \ell_L \equiv (\nu_L, e_L)$ denote  $SU(2)_W$-doublets of 
  left-handed quarks and leptons, respectively. 
  
   It is well known that  $U(1)_{B+L}$  renders the electroweak $\theta_{\rm w}$ unphysical \cite{Anselm:1992yz, Anselm:1993uj,Anselm:1990uy}. 
     This can be seen from the fact that, due to 
  anomalous non-conservation of $B+L$ current 
      \begin{equation}\label{JAnom}
    \partial^{\mu} J_{\mu}^{(B+L)} \propto  W\tilde{W} \,
    \end{equation} 
 under the  $B+L$ -transformation (\ref{B+L}),   the $\theta_{\rm w}$-term shifts, 
      \begin{equation}
      \label{Tshift}      
  \theta_{\rm w} \, \rightarrow \, \theta_{\rm w} \, +N_f\, \alpha \,,
      \end{equation}  
          where $N_f$ is a number of flavours in the Standard Model. 
          
 Alternatively, the vanishing of TSV  can be understood from the fact that the fermions deposit zero modes in the instanton background. 
     
    \subsection{Emergence of $\eta_{\rm w}$} 
    
     We thus observe the following scenario. 
    In a fermion-deprived version of the  Standard Model, the TSV is non-zero. This implies the presence of a massless pole
    in the Chern-Simons $3$-form.  Once the fermions are introduced,  the TSV vanishes. 
   This implies that the pole gets shifted to a non-zero value 
   $p^2 = m_{\eta}^2$.  
    A non-zero pole in the $3$-form correlator corresponds to  
    a pseudo-scalar particle, since   
   a massive $3$-form propagates exactly one physical degree of freedom.   We are thus led to the conclusion that the electroweak sector must contain a new particle, 
  $\eta_{\rm w}$-meson.

    In order to make this connection explicit at the level 
    of the effective theory, let us first outline the general consistency arguments
   \cite{Dvali:2005an, Dvali:2005ws}  indicating that the removal of TSV must be accompanied by the emergence of a pseudo-scalar. 
     Although this part of the story is rather general and applicable to an arbitrary gauge sector with 
   $\theta$-vacua, we shall use notations specific to the electroweak theory, which is our main target. 
   
    First,  following  \cite{Dvali:2005an}, we argue that the existence of a massless pole in TSV fixes the low-energy EFT in the following way, 
  \begin{equation} \label{KEFT}
  L = \Lambda^4 {\mathcal K(E/\Lambda^2)}\,  + \,...\,.
  \end{equation}
   Here,   ${\mathcal K(E/\Lambda^2)}$ is an even algebraic function of its argument, and the ellipses stand for all possible high-derivative terms. 
    Due to the periodicity of physics in 
 $\theta_{\rm w}$,  the first derivative of ${\mathcal K }$, 
 must be an inverse 
 of the periodic function.

      When applied to electroweak 
      $\theta_{\rm w}$, the equation (\ref{KEFT}) gives the first surprise. 
    In order to match both the proper periodicity in $\theta_{\rm w}$,  as well as the TSV, the scale,  $\Lambda$,  
 must be set by a proper power of the instanton rate. 
 Therefore, it is exponentially suppressed relative to the scale $v$ below which the instantons are constrained. 
  At the same time, the scale $v$ sets the perturbative mass gap in the theory. 
  
    Thus, the equation (\ref{KEFT}) creates an impression that  the effective cutoff of EFT of the $3$-form is well below the gap. The resolution of this seeming puzzle is that the massless  $3$-form contains no propagating degrees of freedom. Its sole contribution to physics is the existence of a constant 
   ``electric"  field, 
  $E$,   which parameterizes different vacuum states.  
    
   Notice that  for a constant $E$, all high-derivative 
   terms drop out of the equation of motion, which takes the 
   form, 
   \begin{equation} \label{EqE} 
   \partial_{\mu} \left (\frac{\partial {\mathcal K(E/\Lambda^2)}}{\partial E} \right ) \, = \,  0 \,, 
   \end{equation} 
   and is solved by an arbitrary constant $E$.

    In the absence of fermions, the theory contains no massless mobile sources for 
    $E$.   Correspondingly,  each value of $E$ marks an exact vacuum state. 
    The Hilbert space of the theory thus splits into 
   an infinite set of super-selection sectors parameterized 
 by $E$. These are the $\theta_{\rm w}$-vacua,  
  of the electroweak theory described in the $3$-form language.  This description is exact.  The correspondence 
  between the two parameterizations (for a small angle) is $\theta_{\rm w} \sim E/\Lambda^2$. 
   
 In order to eliminate this vacuum structure,  one  must remove the massless pole from  the correlator (\ref{CCcorr}). That is, one must put the 
 $3$-form in the corresponding ``Higgs"  phase.
  The massive $3$-form propagates a single pseudo-scalar 
  degree of freedom. Correspondingly,  the necessary 
  condition is the existence of a pseudo-scalar $\eta_{\rm w}$ such that, up to a boundary term, the theory in invariant under the following shift symmetry
    \begin{equation} \label{shift}
  \eta_{\rm w}  \rightarrow \eta_{\rm w} + {\rm const.} \,. 
   \end{equation}
  Then, modulo the irrelevant high derivative 
   terms,  the  
   Lagrangian in uniquely fixed as 
     \begin{equation} \label{KetaEFT}
  L = \Lambda^4 {\mathcal K(E/\Lambda^2)}\,  + 
  \frac12 (\partial \eta_{\rm w})^2  +  \frac{\eta_{\rm w}}{f_{\eta}}\Lambda^2 E \,,    
  \end{equation}
  where $f_{\eta}$ is a scale. 
 It is easy to see that, after integrating out  the $3$-form, the theory reduces to a theory of a pseudo-scalar, 
      \begin{equation} 
  L\, = \, 
  \frac12 (\partial \eta_{\rm w})^2  
  - V(\eta_{\rm w}) \,,    
  \end{equation}
 with the potential satisfying the relation, 
       \begin{equation} \label{Vder}
  \frac{\partial V(\eta_{\rm w})}{\partial \eta_{\rm w}}
  \propto E (\eta_{\rm w}) \,,
  \end{equation}
 where $E (\eta_{\rm w})$ as a function of
 $\eta_{\rm w}$ is the solution of the following algebraic equation:
       \begin{equation} \label{Eeta}
      \frac{\partial {\mathcal K}}{\partial E}  \propto 
      \frac{ \eta_{\rm w}}{f_{\eta}} \, + \, \theta_{\rm w} \,.   
  \end{equation}
   Notice that in this equation, $\theta_{\rm w}$-angle  
  enters as an arbitrary integration constant. 
   It is obvious that the $\eta_{\rm w}$-boson makes 
 $\theta_{\rm w}$ unobservable. 
   It suffices to notice that, as indicated by the equation (\ref{Vder}),  at any extremum of the potential,  the $CP$-violating electric field $E$ is zero.

    Next, as shown in \cite{Dvali:2005ws},  whenever  there exist an anomalous current $J_{\mu}$ satisfying
   \begin{equation}  \label{anomEJ}
   \partial^{\mu} J_{\mu} =\Lambda^2 E  \,, 
  \end{equation}      
the generation of the mass gap can be deduced without an explicit reference to an elementary 
$\eta_{\rm w}$.   Indeed, including in the effective 
action the anomalous coupling, 
     \begin{equation} \label{EJcoupling}
  L = \Lambda^4 {\mathcal K(E/\Lambda^2)}\,  + 
    \frac{\Lambda^2}{f_\eta^2}E \frac{1}{\Box} \partial^{\mu} J_{\mu} \,,    
  \end{equation}
  and taking into account the anomalous divergence (\ref{anomEJ}), it is 
 easy to see that $E$ propagates a massive pseudo-scalar. 
    
   Finally, it has been suggested in \cite{Dvali:2013cpa, Dvali:2016uhn} and systematically argumented in 
   \cite{Dvali:2017mpy} that, if  $J_{\mu}$ represents a fermionic Noether current  of some anomalous $U(1)$-symmetry, the 
  story reduces to (\ref{KetaEFT}) with 
   the pseudo-scalar $\eta_{\rm w}$ emerging as the 
   pseudo-Goldstone phase of the 
 fermion condensate of the corresponding 
 't Hooft determinant. In other words, very general consistency arguments indicate that 
 whenever anomalous $U(1)$ removes TSV, the corresponding 
 't Hooft determinant must condense and its pseudo-Goldstone phase must dynamically relax the $\theta$-term. 
  
   As already noticed in \cite{Dvali:2005ws}, a classic 
   realisation of the above scenario is the 
   elimination of $\theta_{QCD}$ by the chiral symmetry 
   of a massless quark. The pseudo-Goldstone degree 
   of freedom in this case is the phase of the chiral quark 
   condensate described by the $\eta'$-meson.

   Applying the same general reasoning to the electroweak sector of the SM, we are lead to a conclusion that on top of the $B+L$-violating 't Hooft vertex, there must  exist 
   its condensate. The elimination 
  of $\theta_{\rm w}$ then can be understood as its dynamical 
  relaxation by the phase of the condensate, 
  described by  the $\eta_{\rm w}$-meson. 
     
   In  the subsequent chapters, we shall support 
   the existence of $B+L$ - violating condensate  
  by an explicit standard instanton calculation
   as well as by 
   a new approach of resolving the instanton
   contribution in terms of multi-particle amplitudes.

\section{Fermion Condensate}

\begin{figure}[!ht] 
\centering
\includegraphics[width=.5\linewidth]{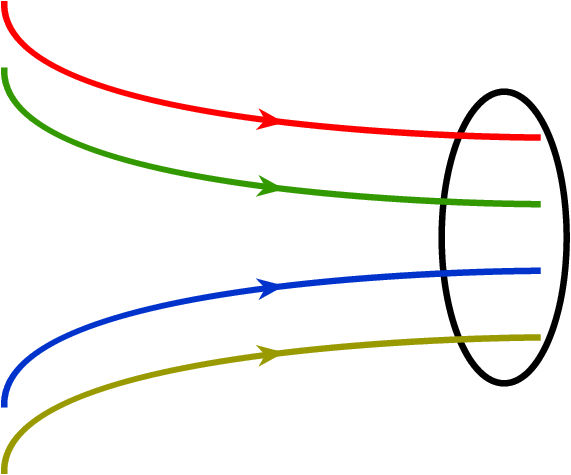}
\caption{Graphical representation of the t'Hooft vertex.}
\label{VVV}
\end{figure}

 We shall now illustrate the existence of a fermion condensate by an explicit computation
 extending the previous analysis \cite{Dvali:2024zpc} \footnote{More extended analysis shall be given in \cite{condensate}.}.  
The instanton index is correlated with the change in the $B+L$ number via the anomaly. The index is given by the equation,
\begin{equation}
\Delta Q_{B+L} = N_f \nu~,
\end{equation}
where $Q_{B+L}$ is the charge corresponding to the $J_{B+L}$ current, $\nu$ is the instanton index, and $N_f$ is the number of Standard Model fermion generations. Without loss of generality, let us consider $\nu = 1$ and $N_f = 1$. The anomaly equation relates $B+L$ violating processes to instanton transition probabilities, which scale as $\sim e^{-\frac{2\pi}{\alpha}}$. Thus, the t'Hooft vertex is generated:
\begin{equation}
\mathcal{L} \sim e^{-\frac{2\pi}{\alpha}} qqql.
\end{equation}
  The diagrammatic representation 
  of the t'Hooft vertex (see Fig.\ref{VVV})
  is given by a merger of all $SU(2)$-doublet fermion lines. 

 We shall later argue that the existence of the t'Hooft vertex implies the existence of the condensate, and vice versa.   
 Since in the electroweak theory, the t'Hooft vertex is a well-understood construct, the above relation provides a shortcut argument for the presence of the condensate.  
 
 However, we shall first compute the condensate via the standard instanton calculus, and later 
 rethink it in the diagrammatic language
 that shall make the connection with the 't Hooft 
 vertex is very transparent.

\subsection{Fermion condensate from instantons} 

In the Standard Model, a unit instanton has $2N_f$ zero modes. Therefore, the instanton measure \cite{Affleck:1980mp} takes the form:
\begin{eqnarray}
\left(\frac{2\pi}{\alpha_2(v)}\right)^4 \int d^4z \int \frac{d\rho}{\rho^5}~e^{-\frac{2\pi}{\alpha_2(\rho)} - 2\pi^2 v^2 \rho^2} (\mu \rho)^{2N_f},
\end{eqnarray}
where $\mu$ is an IR regulator, introduced in the Lagrangian via the terms  $\mu qq$ and $\mu ql$. At the end of the calculation, it should be set to zero. 

In the instanton background, the propagator is expressed via the zero modes ($\Psi$):
\begin{equation}
\bra{x} \frac{1}{\hat{\cal D}+i\mu} \ket{y} = \frac{\Psi_0^{\dagger}(x-z) \Psi_0(y-z)}{i\mu} + \Delta (x-y) + \mathcal{O}(\mu)~.
\end{equation}

\begin{figure}[!ht]
\centering
\includegraphics[width=.5\linewidth]{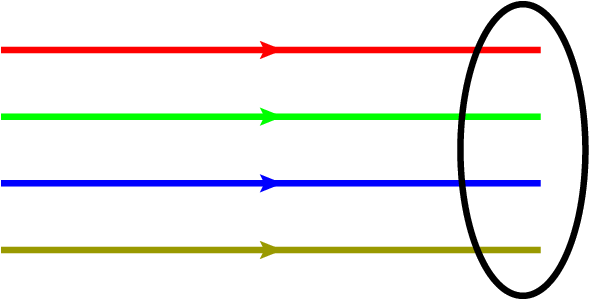}
\caption{Graphical representation of the Condesate.}
\label{condesate_thooft}
\end{figure}

Using the expression above, we can compute the t'Hooft vertex by inserting the propagators at different points, or compute the condensate by considering all external legs inserted at the same point, see the figure \ref{condesate_thooft}. 

Thus, we get:
\begin{eqnarray} \label{q12}
\langle (qqql)^{N_f} \rangle \sim \left(\frac{2\pi}{\alpha_2(v)}\right)^4 \int d^4z \int \frac{d\rho}{\rho^5}~e^{-\frac{2\pi}{\alpha_2(\rho)} - 2\pi^2 v^2 \rho^2}  \left( \bra{0} \frac{\mu \rho}{\hat{\cal D}+i\mu} \ket{0} \right)^{2N_f}.
\end{eqnarray}

We can integrate this expression. Since for the Yukawa couplings $y \ll 1$, we can ignore the Dirac masses. Then, the only scale the integral can produce is $\frac{1}{\rho}$. Therefore, we obtain:
\begin{eqnarray} \label{q12cond}
\langle (qqql)^{N_f} \rangle \sim \left(\frac{2\pi}{\alpha_2(v)}\right)^4 e^{-\frac{2\pi}{\alpha_2(v)}} v^{6N_f} \int \frac{d(\rho v)}{\rho v}~e^{-2\pi^2 v^2 \rho^2}  (\rho v)^{\frac{43 - 8N_f}{6}} (\rho v)^{-6N_f}\times C(N_f),
\end{eqnarray}
where 
\begin{eqnarray}
    C(N_f)=(-1)^{N_f}\left(\frac{8}{\pi^2}\right)^{2N_f-1}\frac{1}{(3N_f-1)(6N_f-1)}\,.
\end{eqnarray}
This expression is convergent for the toy model with $N_f = \frac{1}{2}$, but is divergent for realistic models with more 
flavors. 
  This divergence creates a (false) impression that for a large number of fermion flavors the 
  fermion condensate is sensitive to the UV-cutoff of 
  the SM.  Already the following common sense  EFT argument tells us that  this cannot be the case. 
 
 Indeed, the sources of the fermion condensate are the electroweak 
instanton effects.   These instantons are constrained in IR 
by the length  $\rho  \sim v^{-1}$.   Correspondingly, the 
dominant ones are the instantons of this size, since 
they possess the smallest Euclidean actions. 
 For more compact instantons, $\rho  \ll v^{-1}$,
  the action 
 is suppressed due to the lower value of the 
 running gauge coupling $\alpha_2(\rho)$ at the corresponding scale.  

 With this understanding, it looks natural that the physical 
 effects of instantons cannot be exponentially UV-sensitive in the number of condensed fermion flavors. 
  One argument supporting this view comes from 
  the analytic continuation.  

 A more precise form of the integral is,
\begin{eqnarray}
\langle (qqql)^{N_f} \rangle \sim \left(\frac{2\pi}{\alpha_2(v)}\right)^4 e^{-\frac{2\pi}{\alpha_2(v)}} v^{6N_f} \times C(N_f)\times \frac{1}{2}{(2\pi^2)^{\frac{44N_f-43}{12}}}\Gamma \left(\frac{43-44N_f}{12}\right).
\label{qS12}
\end{eqnarray}
Analytically continuing the above expression, we get,   
\begin{eqnarray} 
\langle (qqql)^{N_f} \rangle \sim \left(\frac{2\pi}{\alpha_2(v)}\right)^4 e^{-\frac{2\pi}{\alpha_2(v)}} v^{6N_f} \times F(N_f),
\end{eqnarray}
where
\begin{equation}
    F(N_f)=\frac{(-1)^{N_f}}{32 (2\pi^2)^{\frac{31}{12}}}\frac{2^{8N_f}(2\pi^2)^{\frac{5N_f}{3}}}{(6N_f-1)(3N_f-1)}\Gamma\left(\frac{43-44N_f}{12}\right)\,,
\end{equation}

\[
\begin{array}{c|c}
N_f & |F(N_f)| \\
\hline
\frac{1}{2} & 0.00248592 \\
1           & 0.658208 \\
2           & 93.4076 \\
3           & 1433.89 \\
4           & 25610.4 \\
\end{array}
\]

 where the table indicates absolute values of the function 
 $F(N_f)$ for different numbers of flavors.

   In the next section, we shall support the statement 
   about the UV-insensitivity of the condensate from a different angle. 
  Namely, we shall argue that the effect is of the same nature as the seeming divergence of multi-particle amplitudes with the particle number. This divergence 
  is unphysical and is removable by a proper re-summation.   

Before concluding this section, we wish to clarify a potentially puzzling point concerning the fermion condensate in the effective low-energy theory. At first glance, one might expect that integrating out heavy fermions (such as the top quark) would allow instantons to generate ’t Hooft vertices—and hence condensates—with fewer fermion fields. This, however, would appear to contradict the topological index theorem, suggesting the possibility of non-perturbative processes within the SM that violate $B+L$ by other than $3\nu$ units. A more careful analysis shows that this does not occur: once heavy fermion fields are integrated out, no new ’t Hooft vertices or associated fermion condensates emerge at low energies, because the action of the electroweak constrained instantons diverges. For further technical discussion, see, e.g., \cite{Georgi:1993dy}

 \subsection{Resolving instantons as multi-$W$ processes}
 
 In this chapter, we shall argue for the UV-insensitivity 
 of the fermion condensate from 
 the point of view of resolving the instantons as multi-particle 
 processes. 
  Our reasoning incorporates the following 
  two points: 

\begin{itemize} 
  
  \item 
 The resolution of a non-perturbative instanton process in terms of a multi-particle amplitude  
 \cite{Dvali:2022vzz};
 
 \item  The bound on the enhancement of the process due to the microstate degeneracy \cite{Dvali:2019ulr, Dvali:2020wqi}. 

  \end{itemize}

   The first  argument that we shall use is a part of a wider program in which the non-perturbative effects are understood as the perturbative multi-particle  processes
(see, \cite{Dvali:2022vzz} and references therein).     
  This correspondence has been explicitly demonstrated 
 by reproducing some known semi-classical 
 results of non-perturbative processes by perturbative computation of multi-particle scattering. 
 
  For example,  in gravity the famous Hawking's black  hole evaporation has been described as a perturbative quantum re-scattering of a large number of gravitons \cite{Dvali:2011aa}.
   Likewise, the semi-classical amplitude of black hole formation in high-energy particle scattering has been reproduced as a quantum process of scattering of two highly energetic gravitons into many soft ones \cite{Dvali:2014ila, Addazi:2016ksu}.  
   Next,  a semi-classical Schwinger effect  has been reproduced as a fully perturbative multi-photon process 
   \cite{Dvali:2022vzz}. Also, an instanton in the $(1+1)$-dimensional Electrodynamics  was resolved via coherent state construction 
\cite{Berezhiani:2024pub}.

\begin{figure}[!ht]
\centering
\includegraphics[width=0.5\linewidth]{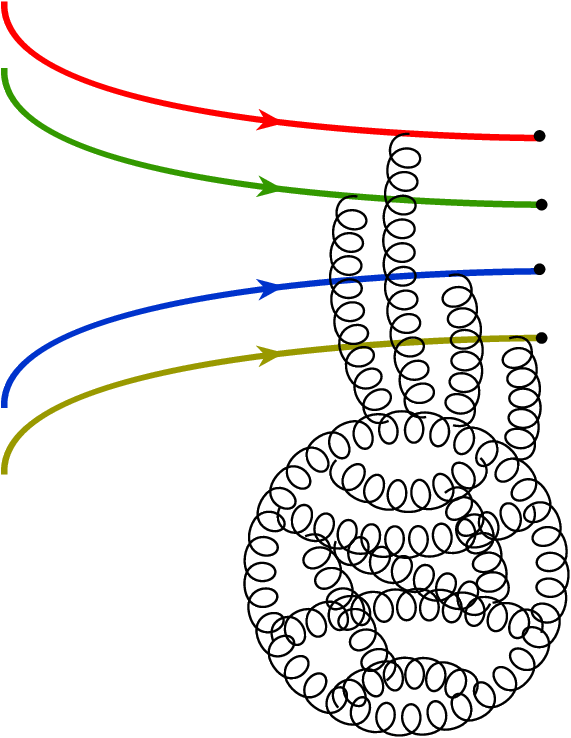}
\caption{Representation of the instanton part of the condensate via $W$-bosons.}
\label{InstantonD}
\end{figure}

    Based on the above,  we adopt the map between a non-perturbative vacuum instanton process and a perturbative multi-particle scattering.   The key is to think about the semi-classical 
  instanton transition as a quantum transition process with a high number of intermediate quanta ( see a schematic 
 diagram Fig. \ref{InstantonD}). 
  This correspondence allows to view the instanton process 
  diagrammatically,  as a vacuum transition process with a large number of intermediate $W$-bosons.  Regardless of the technical difficulty of the actual computation,  this gives a powerful handle on the problem since it allows 
  us to disentangle the real physical effects from 
  fictional effects emerging from the breakdown of perturbation theory. 
  
   The second part of the story is the 
 upper bound on the microstate degeneracy of the instanton. 
 It has been argued \cite{Dvali:2019jjw, Dvali:2019ulr, Dvali:2020wqi} that the microstate entropy of 
any physical entity (a state or a process) with a localization  length-scale $\rho$, is bounded by, 
\begin{equation} \label{Smax}
S_{max} \sim \frac{1}{\alpha (\rho)} \,, 
\end{equation}
where $\alpha (\rho)$ is the relevant running coupling 
evaluated at the scale $\rho$.   
In \cite{Dvali:2020wqi}, the entities saturating this bound were called ``saturons".  
Most importantly for the present discussion, the above bound constrains the instantons 
\cite{Dvali:2019ulr}. 
  The instanton degeneracy factor cannot 
  exceed  ${\rm e}^{S_{max}}$ without invalidating the
  EFT description. In the latter case, the theory must change the regime. 
  
  A particular source of the microstate degeneracy 
  is provided by the fermionic zero modes.  In the present case their number is given by $2N_f$. 
  The saturation criterion, therefore, restricts the number 
  of instanton fermionic zero modes as $2N_f <   2\pi/\alpha$ \cite{Dvali:2019ulr}. 
  Physically, this bound is rather transparent, since 
  the presence of fermion zero modes enhances the 
  instanton contribution into a process by a factor e$^{2N_f}$,  
 thereby,  promoting the point of saturation into an unsuppressed instanton process.   That is, 
 at the saturation point, the collective ('t Hooft) coupling 
 $\alpha N_f$ becomes critical, and instanton rates 
 are unsuppressed.   Beyond this point, the would-be-strong instanton effects invalidate the description, and the theory must change the regime. 

Notice that this saturation is correlated with the breakdown of the fermionic loop expansion, which 
contribute as the power series in the flavor 't Hooft coupling  $\alpha N_f$ \cite{Dvali:2019ulr}.

 This is not the case for electroweak instantons. 
   Indeed,  given the number of $SU(2)_W$ fermion flavors in the SM, $N_f = 3$,  the electroweak instantons are 
     well below the saturation. Correspondingly, no change 
     of the regime of the theory is expected by the instantons.  This is a fully non-perturbative statement.

\begin{figure}[!ht]
    \centering
    \includegraphics[width=0.5\linewidth]{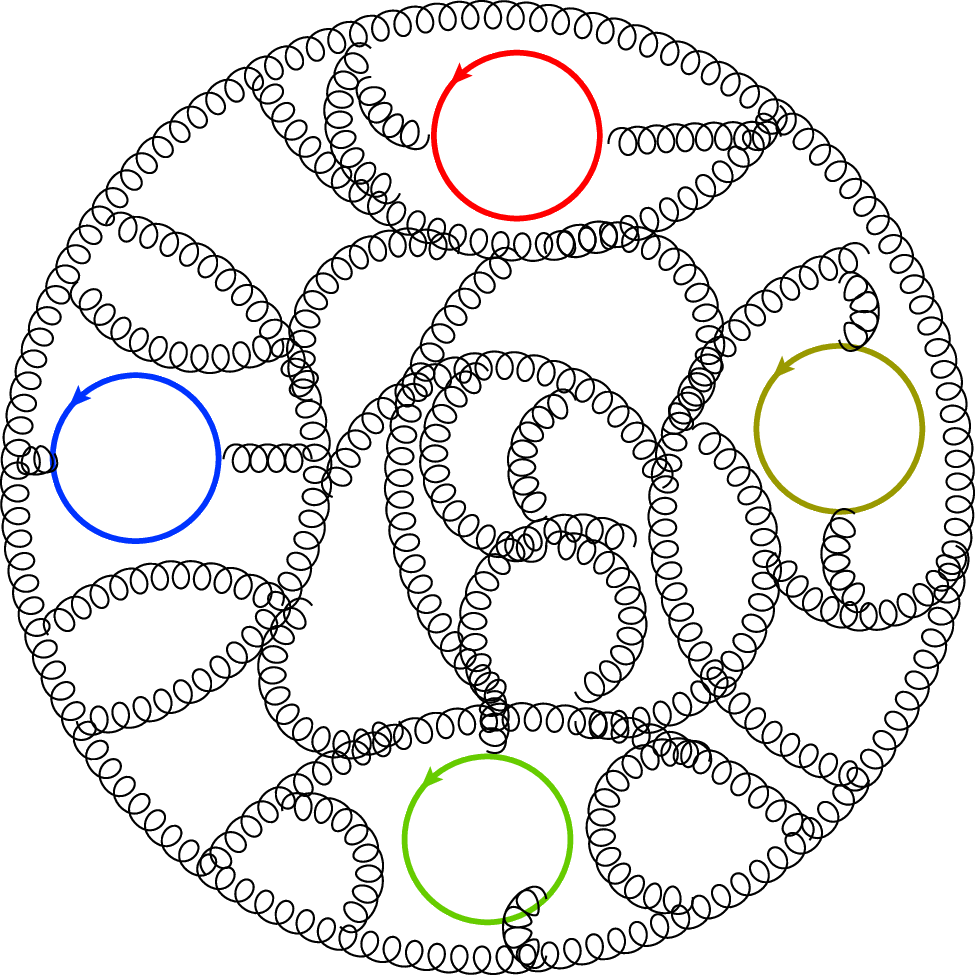}
    \caption{ An example of a multi-particle process, with many virtual $W$-bosons 
   and intermediate fermion loops.  
    Even for an under-saturated value of the 't Hooft coupling, $\alpha N \ll 1$, the multiplicity  of diagrams grows factorially with the number of participating quanta $n$.  
   However, this enhancement 
   signals the breakdown of perturbation theory 
   and cannot be trusted beyond the point of optimal
   truncation $n \sim 1/\alpha$.}
    \label{Fermion loop}
\end{figure}

   Now, naively, even for undesaturated 
   collective coupling, $\alpha N_f \ll 1$, 
   for the processes with a large number of participating particles $n \gg 1$ 
   (see, e.g., Fig.
   \ref{Fermion loop}),   the combinatorics  grows factorially with  $n$ leading to 
  a fictional enhancement of the effect. 
  This enhancement is, however, related to the breakdown of perturbation theory rather than with 
  a real physical enhancement of the effect.  

  Such divergences cannot be trusted
  beyond the point of the optimal truncation. 
 This point is reached when the number of quanta $n$ 
 participating in the process reaches the critical value   \cite{Dvali:2020wqi}, 
 \begin{equation}\label{optP}
 n \simeq  \frac{1}{\alpha}\,. 
\end{equation} 
 Following the above paper, the origin of the optimal truncation point can be understood by the following counting. 
 Consider a process that involves a transition 
 between a few (say two) to $n \gg 1$ quanta. 
 For example, the $n$ participating quanta can appear 
 as on-shell final (or initial) state particles 
 in $2\rightarrow n$ tree-level scattering process, or 
 as a virtual intermediate quanta of $n$-loop diagram 
 describing the $2\rightarrow 2$ transition. 
 In the latter case, the $2\rightarrow n$ transition rate can be extracted as the imaginary part of the amplitude. 
  
  This rate scales as 
  $\sim n!\alpha^n$, where the factorial enhancement comes from the combinatorics of  
  diagrams and the suppression comes from $n$ interaction vertexes. 
   The perturbative expansion breaks down when  
   $n+1$ order effect catches up with the one coming from the order $n$.  
  This happens for (\ref{optP}), which defines the point of optimal truncation. 
 Using Stirling's approximation, it is easy to see that the corresponding 
 transition rate scales as, 
   \begin{equation} \label{expn}
  \Gamma \sim e^{n} \simeq e^{-\frac{1}{\alpha}}\,.
  \end{equation}   
 Beyond this point, the series has to be 
 re-summed with higher-order effects, giving more suppression.  
 The leading order effect is thus given by (\ref{expn}). 
  Thus, the exponential suppression characteristic 
  of non-perturbative semi-classical effects is reproduced by a 
  perturbative multi-particle process.

   Notice that there exist general non-perturbative arguments \cite{Dvali:2020wqi}
    indicating that within the validity of given degrees 
    of freedom, the matrix element of any process of the type few$\rightarrow n$ is bounded from above by  $e^{-n}$
    modulo the microstate degeneracy factor $e^{S}$. The total rate thus scales as,   
  \begin{equation} \label{expnS}
   \Gamma\sim  e^{-\frac{1}{\alpha} + S}\,.
  \end{equation}

  Thus, the non-perturbative physics is telling 
 us that the enhancement factor due to the divergent integral 
 in (\ref{q12}) cannot exceed the  enhancement factor 
 by the instanton's microstate entropy due to the fermionic zero modes. 
  Moreover, the fermion enhancement of the microstate 
  entropy is essentially independent of the instanton scale. 
   We thus reach a  conclusion that 
   the  final enhancement factor due to fermion zero modes cannot exceed the exponential factor in $N_f$.

   A schematic diagrammatic representation of the process 
   responsible for the creation of the condensate as 
   well as of the 't Hooft vertex is given
  in Fig. \ref{InstantonD}.  The solid colored lines represent different flavors of the fermion doublets, whereas the wavy lines describe the virtual $W$-bosons.   The multi-loop blob
  of virtual $W$-bosons describes the instanton process.  This instanton represents a process 
  with the participation of $n \simeq 2\pi/\alpha$
  coherent $W$-bosons with characteristic momentum 
  $1/\rho$.  The emergence of the zero modes 
 that are the source of the 't Hooft vertex 
 as well as of the fermion condensate 
 is due to the connection between the instanton 
 blob and each fermion line via a virtual 
 $W$ lines.   To the leading order, the number of connector  $W$-lines is the same as the number of fermion doublets, $3N_f$.

   Now increasing the virtual momenta 
  in the connector $W$-lines, can increase the 
  constituency of the instanton blob by 
  populating it with a larger number of 
  virtual quanta, $n \gg  2\pi/\alpha$,
  including the additional fermionic loops (see, Fig \ref{Fermion loop}). 
  As we already discussed, this leads to a fictional 
  factorial growth of the contribution, which, in reality, signals the breakdown of 
  perturbation theory and goes away after the proper re-summation. 
 The enhancement cannot exceed an exponential factor in $N_f$.

This settles the issue of cutoff-(in)sensitivity of the fermion condensate.   In a full-fledged calculation, the integral over $\rho$ in (\ref{q12}) must be effectively cutoff around the scale $v$. The enhancement factor emerges after the re-summation beyond the point of optimal truncation and boils down to an exponential factor in $N_f$.

  The diagrammatic visualization of the instanton process allows us to establish an immediate one-to-one correspondence between the generation of 
  the vertex of the 't Hooft determinant and 
its expectation value.  Indeed, in the diagrammatic language 
the 't Hooft vertex absorbs $3N_f$ fermion lines.
 Since the  Lagrangian possesses no flavor-changing 
 vertexes, the fermion lines can only end on the condensate.

  \section{Where is $\eta_{\rm w}$?}  
 
  In understanding the features of  $\eta_{\rm w}$, it is illuminating to  confront it with 
  the situation with the $\eta'$-meson of QCD in case 
  of a massless (or a light) quark.   It was already 
  pointed out in \cite{Anselm:1993uj} that the way 
  $B+L$ makes  $\theta_{\rm w}$ unphysical is
  very similar to how the axial  $U(1)_A$-symmetry of a massless quark eliminates $\theta_{QCD}$.  
    However, another important part of the 
    parallel has been overlooked. 
    
    First, in QCD, there exists an associated 
    degree of freedom in the form of the $\eta'$-meson
  which gets rid of the massless pole in TSV and simultaneously relaxes  $\theta_{QCD}$ to zero. 
     This degree of freedom emerges as the phase 
     of the fermion condensate of the 't Hooft determinant.  
 A small explicit breaking of $U(1)_A$-symmetry by the quark mass makes the 
 cancellation of $\theta_{QCD}$ imprecise, but does not 
 jeopardize the existence of the $\eta'$-meson.

 The point made by us in \cite{Dvali:2024zpc} and further justified 
 in the present work, it is that there must exist an analogous boson in the electroweak case. 
       The following parallels exist between the two systems. 
       
        In both cases, 
     in the pure gauge sector without fermions 
     the vacuum has a topological structure. Instantons
     induce a non-zero TSV and correspondingly produce  
   the physical  $\theta$-vacua. In both theories, upon introduction of (good quality) anomalous symmetries  
   acting on fermions ($U(1)_A$ in QCD and $U(1)_{B+L}$ in 
   electroweak theory), the respective $\theta$-terms become unphysical.
   This is in full agreement with the emergence of the fermionic zero modes that, on one hand, suppress the instanton transitions and, on the other hand, lead to the generation of the fermionic 't Hooft determinant vertexes. These vertexes 
 break the respective anomalous global symmetries explicitly down to their anomaly-free discrete subgroups.

 In addition, in both cases, the 
   corresponding 't Hooft determinants acquire the vacuum expectation values, breaking the respective 
   anomalous symmetries also spontaneously.  This was 
   well-known for QCD.  We have argued that the similar effect 
   takes place in the electroweak theory. 
 The condensation of the fermion determinant is in accordance with the previously-suggested general  connection between the elimination of topological susceptibility by anomalous symmetry and the existence of a fermion condensate that breaks the same symmetry spontaneously \cite{Dvali:2017mpy}.

   Now, in QCD, the fluctuation of the phase of the 
   condensate of the fermion determinant  is 
   a physical pseudo-Goldstone degree of freedom, 
   $\eta'$. The vanishing of the $\theta$-terms can be understood as the dynamical relaxation of the pseudo-Goldstone. However, this degree of freedom is also necessary for explaining the removal of the massless pole in TSV. This is the source of the mass 
   of $\eta'$ \cite{Witten:1979vv, Veneziano:1979ec}.   As already discussed, the phenomenon of $\eta'$ mass-generation from TSV represents a $3$-form analog of the ordinary Higgs effect: $\eta'$ is ``eaten-up" by the Chern-Simons 
   $3$-form, thereby, shifting the pole to a massive value \cite{Dvali:2005an}. 
  In other words, the presence of a pseudo-scalar degree of freedom is the only gauge invariant way for making a $3$-form massive.  

  Our point is that the presence of an analogous degree of freedom is necessary in order to explain the removal of the electroweak TSV by the anomalous $B+L$-symmetry. 
 However, the question is about its origin. 
 
   Of course, there exists an option that  the $\eta_{\rm w}$-particle
is an ``external" degree of freedom accompanying the SM by consistency.  
 However, the existence of the fermion condensate 
 with the right quantum numbers suggests that, 
instead of seeking external help, the SM  itself 
generates the right degree of freedom dynamically.  
This indicates that $\eta_{\rm w}$ can emerge as the collective excitation describing the fluctuation of the phase of the 
fermion condensate.    Such a scenario would be in close correspondence with the story of $\eta'$ in QCD. The similarities between the above theories can be summarized in the table \ref{2sectors}.
  \begin{table}[!ht]
      \centering
      \begin{tabular}{l|c|c}
         Theory: & QCD &  Electroweak  \\\hline
TSV in pure gauge sector & ~~&\\
(physical $\theta$): &   Physical $\theta_{QCD}$ & Physical $\theta_{\rm w}$  \\\hline
TSV removed by anomalous  & ~~~ & ~~~\\  symmetry ($\theta$ rendered unphysical): &  $U(1)_A$& $U(1)_{B+L}$\\\hline
Fermion condensate (flavour-$\det$):&  $\langle \det_{QCD}\bar{q}q \rangle$ & $\langle\det_{Weak}qqql\rangle$  \\\hline
pseudo-Goldstone: &   $\eta'$&    $\eta_{\rm w}$ \\
      \end{tabular}
  \caption{The parallels between QCD and Electroweak
  theories}\label{2sectors}
  \end{table}

  However, there exists certain important differences between the 
  two theories. Taking these differences into account
  leaves some open questions about the features 
  of $\eta_{\rm w}$ particle.  Let us briefly outline the puzzle.  
    
    The drastic difference between the two scenarios 
    is that in QCD, due to the confinement,  quarks
    and mesons represent the valid degrees of freedom in 
    the separate domains of the description. 
    In particular, quarks cannot coexist with the 
    $\eta'$-meson.  This is not  {\it a priori} the case in the electroweak 
    sector. From the first glance, it appears that, without further precautions, 
   the emergence of the $\eta_{\rm w}$-meson 
    does not necessarily invalidate fermions as the acceptable  asymptotic states \footnote{  
    Of course, due to QCD dynamics, the quarks are confined. However,  this effect is independent from the electroweak instantons that form the fermion condensate and produce the $\eta_{\rm w}$-meson.}. 
    
    In such a case the $\eta_{\rm w}$-meson and fermions 
 would coexist within the same energy domain. 
  Notice that, since the condensate is formed by the effects operating around the scale $v$,  the $\eta_{\rm w}$-meson
  must emerge as the collective excitation 
at the same scale.  At the same time the fermions can be 
taken to be much lighter (as this is the case for all fermions in the SM except the top). 
 Thus,   $\eta_{\rm w}$ and fermions can have 
 the overlapping domains of validity of their EFT descriptions. 
  
   In other words, the dramatic distinction between the two cases ($\eta'$ versus $\eta_{\rm w}$)  is the hierarchy of 
   underlying scales.  In QCD, the scale of the 
   condensate formation (the QCD scale) and the value of the condensate are roughly of the same order\footnote{ 
  In QCD with a large number of colors, $N_c$, the value of the condensate grows with $N_c$ and thus is even higher than the QCD scale.}. 
  That is, the scale of physics that creates the condensate
  is not hierarchically larger than the scale of the condensate.  
     The unusual thing about the electroweak case is that the two scales are exponentially separated.  That is, 
     the instanton physics that forms the condensate 
     operates at the scale $v$, whereas the value 
     of the resulting condensate is exponentially smaller.      
     This raises the question about the domain of validity of the  EFT of the $\eta_{\rm w}$-meson. 
     
      For  sharpening the 
     question,  let us choose the fermion Yukawa couplings
   in such a way that the fermion masses, $m_f$, are 
   well below the instanton scale $v$, but are much 
   above the scale of the topological suscebility    $\Lambda$ \eqref{KEFT}:        
     \begin{equation}  \label{hierarchy}
        v \gg m_f \gg \Lambda \,.
     \end{equation} 

     The hierarchy has the following form,

\begin{center}
    
\begin{tikzpicture}[scale=1]
\draw[thick] (0,0) -- (0,5);

\draw[fill=black] (0,0.3) circle (2pt) node[right=8pt] {\( \Lambda \)};
\draw[fill=black] (0,2.5) circle (2pt) node[right=8pt] {\( m_f \)};
\draw[fill=black] (0,3.7) circle (2pt) node[right=8pt] {\( m_{\rm w} \)};
\draw[fill=black] (0,4.7) circle (2pt) node[right=8pt] {\( v \)};

\node[rotate=90] at (-0.5,2.5) {Energy scale};
\end{tikzpicture}
\end{center}

     Let the  UV-cutoff  (upper bound on the validity scale) of EFT of the  $\eta_{\rm w}$-meson be $f_{\eta}$. 
    Let us ask where is the place of $f_{\eta}$ in the above hierarchy? 
   
   Putting $f_{\eta} \ll m_f$  creates a naive problem 
   of UV-completion of the theory of $\eta_{\rm w}$ above the scale $f_{\eta}$.  Indeed, in such a case there would exist
 an ``unpopulated"  energy gap between the two scales.

 On the other hand, putting 
  $f_{\eta} \gg m_f$  is equally puzzling as this degree of freedom would coexist with elementary fermions within a finite interval of scales
 between $f_{\eta}$ and  $m_f$, and impossible to match anomalies between high and low energy scales.  
 Such an arrangement would also be in tension with the $a$-theorem,  since the number of degrees of freedom would increase below the scale $f_{\eta}$. 
    
     From the above considerations, the minimalistic consistent case emerges to be  $f_{\eta} \sim m_f$. 
    This would avoid both of the above issues since fermions 
    and  $\eta_{\rm w}$  would exist in separate domains of validity, and the theory of $\eta_{\rm w}$ would be UV-completed by 
    fermions. 
    
     However, this still leaves the following question. 
    In our calculation of the condensate, the fermion masses 
    were subdominant.  The scale of the physics that forms the condensate is $v$.   It is therefore unclear why 
   the cutoff of $\eta_{\rm w}$ must be sensitive to the 
   fermion mass scale.   Above the scale $m_f$, the 't Hooft vertex is kinematically accessible, and we could think that all the anomalous processes can be described via fermions. Following that existence of the $\eta_{\rm w}$ in this domain is not possible. So, the condensate should start to evaporate, or at least it should not deliver a Goldstone boson anymore. 
   
   One may argue that within the realistic SM  the fermion masses do not obey  the hierarchy (\ref{hierarchy}),  due to the mass of the top quark, and this may affect the story. 
   However, this does not appear to be a fully satisfactory argument, since we would expect that the SM must stay a consistent theory even under the deformed Yukawa couplings that push the 
 top mass well below $v$.  
  
   In summary, 
   the validity scale $f_{\eta}$ and its 
   connection with the scale of instantons $v$ and 
   the fermion masses $m_f$ remains to be
    understood.     
\subsection{Parameters from first principles}
In principle, it should be possible to compute 
from first principles via the instanton calculus
the mass, the decay constant and the effective cutoff of the $\eta_{\rm w}$. Let us consider an operator $\hat{O}(x)$, which, after a $B+L$ rotation, gives the quark multilinear operator that enters in the 't Hooft vertex; in other words,
\begin{equation}
\delta_{B+L} O = q q q l(x),
\end{equation}
where, for simplicity and without loss of generality, we take $N_f = 1$ (generalization to $N_f > 1$ is straightforward). Then, the following correlator,
\begin{equation}
\mathrm{FT} \bra{} O(x) O(0) \ket{}_p = \frac{\rho(m_{\eta_{\rm w}})}{p^2 - m_\eta^2} + \ldots,
\label{Pcorrelator}
\end{equation}
contains the $\eta_{\rm w}$-meson as the lightest degree of freedom in the spectral representation. The process described by the above correlator requires a $\nu=2$ change in winding, corresponding to its total charge $B+L=2$.  Thus, we can deduce that the spectral weight is proportional to
\begin{equation}
\rho(m^2_{\eta_{\rm w}}) \propto \langle q q q l \rangle^2.
\end{equation}
Since this correlator is represented via the $\eta_{\rm w}$'s propagator, we can extract the mass from it. Using the Witten-Veneziano relation \cite{Witten:1979vv,Veneziano:1979ec}
\[
m_{\eta}^2 f_{\eta}^2 \sim \Lambda^4,
\]
where $\Lambda$ is the EFT scale \eqref{KEFT}, we have the possibility to compute  the decay constant $f_{\eta}$.

Also, the above correlator carries the total $B+L$ charge corresponding to the winding number $\nu = 2$. Therefore, the correlator must be computed via the $\nu = 2$ correlated instantons.  Due to this selection rule, the computation should be well defined (due to the absence of mixing with $\nu=0$). Some of the tools necessary for this computation were developed in the context of QCD in \cite{Pisarski:2019upw}. The correlator should contain information about the fermion masses. Unlike the computation of the condensate, this correlator is highly sensitive to the form of the zero modes, since the correlation is given by their overlap on different instanton centers (see the non-local part of the computation from \cite{Pisarski:2019upw}). In other words, the zero modes for the $\nu = 2$ instanton configuration should be computed while carefully retaining their dependence on the Dirac masses. Profiles for the zero modes in the $\nu = 1$ instanton background in the small-mass approximation were derived in \cite{Krasnikov:1979kz}. Following a similar procedure, one can obtain the mass-dependent zero mode solutions for $\nu = 2$, insert them into the correlator expression in \eqref{Pcorrelator}, then subsequently integrate over the instanton moduli space and perform the Fourier transform. This procedure yields the mass of $\eta_{\rm w}$ as a function of the average instanton separation in the dilute gas approximation, as well as the fermion masses.

Moreover, the structure of the correlator \eqref{Pcorrelator} effectively encodes the cutoff scale of the low-energy effective field theory. This is manifested through the emergence of a branch cut in the correlator. Its scale corresponds to the breakdown of the effective description. This can determine the scale $f_\eta$, the cutoff of the theory, and their relation.

  \section{Outlook} 
 
   In this article, following \cite{Dvali:2024zpc},  we have reiterated that the physics 
 of the electroweak vacuum points towards the 
  existence of a new degree of freedom, the $\eta_{\rm w}$-meson.  This boson is sourced by the anomalous $B+L$-current and gets its mass from the TSV of the electroweak 
  vacuum.   In the previous study, we have also used gravity as a ``spectator" for making certain statements exact. 
  However, even decoupling gravity, the structure 
  of the electroweak vacuum offers non-trivial evidence for 
 the existence of $\eta_{\rm w}$. 
  
   The purpose of the present work was to clearly identify and separate the evidence for $\eta_{\rm w}$ that is provided by the pure standard model without any external help.   
   
  The main guideline is a general criterion consisting of two parts.  First, the removal of TSV implies the existence of a 
  pseudo-scalar particle that gets its mass from shifting the  would-be massless pole in TSV to a non-zero value
  \cite{Dvali:2005an, Dvali:2005ws} .  
  Moreover, if  TSV is removed by an anomalous symmetry acting on fermions, the  corresponding
  fermionic 't Hooft determinant acquires  a non-zero 
  expectation value \cite{Dvali:2013cpa, Dvali:2016uhn, Dvali:2017mpy}.   
  The required pseudo-scalar then emerges as the fluctuation of the phase of the fermion condensate
 (for a detailed unified argument, see  \cite{Dvali:2017mpy}).   
 
   The explicit example of this phenomenon is provided by the $\eta'$-meson in QCD with a massless quark. 
   Another, less known, example \cite{Dvali:2024dlb} is the removal of the
  gravitational TSV,  generated by Eguchi-Hanson instantons \cite{Eguchi:1978xp, Eguchi:1978gw},  by the 
  anomalous symmetry of a spin-$3/2$ fermion (gravitino). 
   According to the index theorem \cite{index}, 
 a spin-$3/2$ particle is the only fermion possessing the zero modes in the Eguchi-Hanson background.  
 The generation of the gravitino condensate by the instanton zero modes has been demonstrated 
 explicitly \cite{Hawking:1978ghb, Konishi:1988mb}.      
    
   In the present paper, following our recent work \cite{Dvali:2024zpc}, we have applied the similar reasoning 
   to the electroweak vacuum.  It is well understood
   \cite{Anselm:1992yz, Anselm:1993uj}     
  that $\theta_{\rm w}$ is rendered unphysical by the 
  $U(1)_{B+L}$-symmetry of the SM.   
   We thereby expect that, by consistency, the theory must contain a pseudo-scalar that 
   removes TSV.  
    We showed that some crucial ingredients supporting the existence of such a scalar are in place. Namely, the instanton zero modes generate the VEV of the    
    't Hooft fermionic determinant (\ref{qS12}).
       
    As an important byproduct, we have clarified the issue of the 
   (seeming)  exponential cutoff-sensitivity to the number of fermion flavors  appearing in the leading order calculation
   from integration over the instanton size (\ref{q12cond}).     
   This sensitivity has caused uncertainties 
  in the previous literature, creating an impression that for a large number of flavors the fermion condensate may be strongly enhanced by the cutoff physics.   
   
    Adopting the reasoning of \cite{Dvali:2020wqi},  
 we have argued that this divergence is an artefact of extending the multiparticle amplitudes (representing the instanton process)  beyond the point of the optimal truncation. In reality, the maximal 
enhancement by the zero-mode degeneracy of the instanton 
action cannot exceed (\ref{expnS}).  Correspondingly,  the UV-sensitivity of the fermion condensate is absent, and the leading order result is given by (\ref{qS12}).  
  
   In addition, by mapping the non-perturbative semi-classical instanton process on a multi-particle transition amplitude,  in the spirit of \cite{Dvali:2022vzz,  Dvali:2020wqi},  we established a one-to-one correspondence between the existence of the 't Hooft vertex and its vacuum expectation value.  
   
   Thus, our analysis demonstrates the existence of the fermionic condensate. 
 This leads us to the following situation.   
 On one hand, the elimination of the TSV by $B+L$-symmetry demands the 
 presence of  the $\eta_{\rm w}$-boson realising the anomalous symmetry non-linearly.  On the other hand, the theory provides a fermion condensate that breaks the same symmetry spontaneously.  
 Given this evidence, it is reasonable to expect that the required boson originates from the phase fluctuation of the fermion condensate. 
 
  Although the story is similar to the emergence 
  of $\eta'$-meson in QCD, there are some important differences which leave some questions about the nature of the $\eta_{\rm w}$-boson open.  The main puzzle is the 
  relation between the  EFT cutoff ($f_{\eta}$) of the theory 
  of $\eta_{\rm w}$ and the masses of fermions. 
  
   The EFT intuition implies that $\eta_{\rm w}$ must emerge at the scale below the masses of (at least some of) the fermions. However, our calculations indicate that in the case of light fermions, the condensate is insensitive to the fermion masses.  This puzzle requires further clarification. It is possible that it gets resolved by taking into account the mass of the top-quark.

   We must stress that, with our current understanding, it remains an option that gravity 
   is  essential for having $\eta_{\rm w}$
 as an elementary degree of freedom, implying that 
 in pure SM the $\theta_{\rm w}$ is neutralized 
 by a broad resonance composed out of SM fermions 
 that cannot be regarded as a well-defined asymptotic state.  This would imply that 
  in the limit of zero gravity, $\rho(0)=0$ in \eqref{CCcorr}.
  Despite the opposite evidence, discussed in the main body of the paper as well as in \cite{Dvali:2024zpc},  we were not able to exclude such an option.  Of course, since gravity 
  is part of nature, one way or the other 
  $\eta_{\rm w}$ is expected to be a physical state.

   Independently whether the $\eta_{\rm w}$-meson is emergent as a collective excitation from SM physics or has to be introduced 
by consistency as a new elementary degree of freedom, it is expected to be an extremely weakly interacting light particle.  This creates a substantial difference with the case of simply unphysical $\theta_{\rm w}$. 
 
  Indeed,  if $U(1)_{B+L}$ could remove $\theta_{\rm w}$ without 
  making it dynamical, no physical effects of $\theta_{\rm w}$
  can  exist in principle.  
 Situation is dramatically different if  $\theta_{\rm w}$ is 
relaxed dynamically through $\eta_{\rm w}$.  In such a case, the off-vacuum configurations with oscillating $\theta_{\rm w}$ must exist.  Obviously, these are physically distinct from the state  of a non-dynamical and constant $\theta_{\rm w} = 0$. 
 
   Moreover,  cosmologically, the states with oscillating 
 $\theta_{\rm w}$ are maximally probable.   
    Indeed, since the relaxation  time of the $\eta_{\rm w}$-boson 
   is extremely long, with a high likelihood, cosmologically, we are expected to live in a background with 
  $\frac{\eta_{\rm w}}{f_\eta} \sim 1$.  Notice that a possible high-dimensional 
  operators generated by physics beyond the SM, 
  can shift neither the minimum of  $\eta_{\rm w}$ nor its mass substantially. \\

 \section*{Note added}

Shortly before submitting this paper, we received 
  a draft by Giacomo Cacciapaglia, Francesco Sannino, and Jessica Turner, in which they identify the electroweak  $\eta_{\rm w}$  
  with a CP-odd combination of the hydrogen  
and anti-hydrogen atoms.  We are grateful to these 
authors for sharing their preliminary results. 

 While looking for the candidates for $\eta_{\rm w}$ among the existing atomic states is a natural minimalistic approach, in light of our analysis, the validity of such identification is not fully clear.  
Putting aside the question of the contribution of the heavy fermions and focusing on a toy version of the SM with a single light generation, the
condensate of 't Hooft determinant
(\ref{q12}), $\langle qqql \rangle$, from which the $\eta_{\rm w}$-meson emerges as a collective excitation, does have the quantum numbers of a proper hydrogen state. In this respect one can say that the author's proposal is aligned with our setup. 

However, it is doubtful whether an on-shell state of hydrogen-anti-hydrogen, even with the correct quantum number, is the right match for a one-particle $\eta_{\rm w}$-state. For example, the mass gap of the state must be exponentially suppressed, as it must be generated 
   by the electroweak instantons. This is not the case for the states constructed out of on-shell (anti)hydrogen. 
   
 Even if a degree of freedom with such a high mass gap can be identified, it is unclear how it can 
 Higgs the $3$-form. The latter then shall remain massless.  Upon taking into account gravity, 
 which excludes the decoupling of the $3$-form, the $\eta_{\rm w}$ must enter as a different degree of freedom which Higgses it. 
 
 To summarize, the role of a pseudo-Goldstone that is required for Higgsing the $3$-form is unlikely to be played by a broad resonance composed out of the heavy modes.  This however does not exclude the mixing between the two types of degrees of freedom (hydrogen-type heavy modes and the ``genuine" $\eta_{\rm w}$) which can lead to potentially-interesting physical effects.

\section*{Acknowledgments}
We are grateful to Lasha Berezhiani and Georgios Karananas for insightful discussions. AK also acknowledges the support received during his sabbatical at the Max Planck Institute for Physics in Munich, where part of this work was carried out. The work of GD was supported in part by the Humboldt Foundation under the Humboldt Professorship Award, by the European Research Council Gravities Horizon Grant AO number: 850 173-6, by the Deutsche Forschungsgemeinschaft (DFG, German Research Foundation) under Germany’s Excellence Strategy - EXC-2111 - 390814868, Germany’s Excellence Strategy under Excellence Cluster Origins EXC 2094 – 390783311. The work of AK and OS was partially supported by the Australian Research Council under the Discovery Projects grants DP210101636 and DP220101721.     \\

\noindent {\bf Disclaimer:} Funded by the European Union. Views and opinions expressed are, however, those of the authors only and do not necessarily reflect those of the European Union or European Research Council. Neither the European Union nor the granting authority can be held responsible for them.\\

\end{document}